\documentclass[11pt,twoside]{article}

\usepackage{asp2006}
\usepackage{epsf}
\usepackage{psfig}
\usepackage{lscape}

\begin{document}
\title{Constraining feedback in galaxy formation: using galaxy and AGN surveys to shed light on "gastrophysics"}
\author{Pierluigi Monaco}
\affil{Dipartimento di Astronomia and INAF-Osservatorio Astronomico di Trieste, via Tiepolo 11, 34143 Trieste}

\begin{abstract}
We present some results of the new {\sc morgana} model for the rise of
galaxies and active nuclei, and show that the improved physical
motivation of the description of star formation and feedback allows to
get hints on the physical processes at play.  We propose that the high
level of turbulence in star-forming bulges is at the base of the
observed downsizing of AGNs.  In this framework it is also possible to
reproduce the recently obtained evidence that most low-redshift
accretion is powered by relatively massive, slowly accreting black
holes.  Besides, we notice that many galaxy formation models
(including {\sc morgana}) fail to reproduce a basic observable, namely
the number density of $10^{11}$ M$_\odot$ galaxies at $z\sim1$, as
traced by the GOODS-MUSIC sample.  This points to a possibly missing
ingredient in the modeling of stellar feedback.
\end{abstract}



\section{Introduction}

Understanding the properties of galaxy and AGN populations is one of
the major challenges of modern cosmology.  However, our poor
understanding of the astrophysical processes that drive the formation
of stars in galaxies has forced cosmologists to use simple
phenomenological recipes to describe these processes.  This is true
both for semi-analytic codes and N-body simulations, for which star
formation and feedback are "sub-grid physics".  As a consequence, once
agreement between model predictions and data is achieved, it is not
straightforward to get hints on the physical processes that have
shaped galaxies.

Besides, more and more information is being gathered on how feedback
from newly formed stars at high redshift regulates star formation by
observing single galaxies or their surrounding intergalactic medium
(see Steidel, these proceedings).  Most importantly, IFU observations
of high-redshift galaxies (Genzel et al. 2006; Swinbank, these
proceedings) are revealing unexpected details that give important and
new clues on the processes at play.

To bridge these sets of data we need a galaxy formation model based on
a more realistic and sophisticated description of feedback and able to
produce predictions for large galaxy datasets.  To this aim we have
developed the galaxy formation code {\sc morgana} (MOdel for the Rise
of Galaxies aNd Active nuclei).

\section{Kinetic feedback in star-forming bulges}

The {\sc morgana} code is described in full detail in Monaco, Fontanot
\& Taffoni (2006).  We show here an example of its ability to obtain
insight on stellar feedback by comparing model predictions on the
evolution of the AGN population to observational data.

Feedback from star formation is implemented using the results of the
multi-phase model of Monaco (2004).  According to this model, the way
feedback works depends sensitively on the surface density and geometry
of the system. In thin and moderately dense systems like spiral discs,
the superbubbles originated by multiple SN explosions blow out of the
system very soon, thus injecting most of their thermal and kinetic
energy to the external halo, while in denser and/or thicker systems
like star-forming spheroids, the energy gets trapped in the
inter-stellar medium.  A consequence of this is a higher level of
turbulence is expected in the cold gas of a star-forming spheroid.
Assuming that the energy injected by SNe is balanced by the loss due
to the decay of turbulence, the velocity dispersion of cold clouds
$\sigma_{\rm cold}$ should scale as:

\begin{equation}
\sigma_{\rm cold}=\sigma_0 \left( \frac{t_\star}{1\ {\rm Gyr}}
\right)^{-1/3}\ {\rm km\ s}^{-1}
\end{equation}

\noindent
The increase of turbulence in star-forming galaxies has been measured,
for instance, by Dib, Bell \& Burkert (2006).  Moreover, star-forming
discs at high redshift are observed to be denser and with a much
higher degree of turbulence (Genzel et al. 2006).  A high level of
turbulence increases the probability that some clouds have high enough
kinetic energy to be ejected out of the galaxy.  As a consequence, gas
will be easily evacuated from small, gas-rich bulges.

\begin{figure}
\plottwo{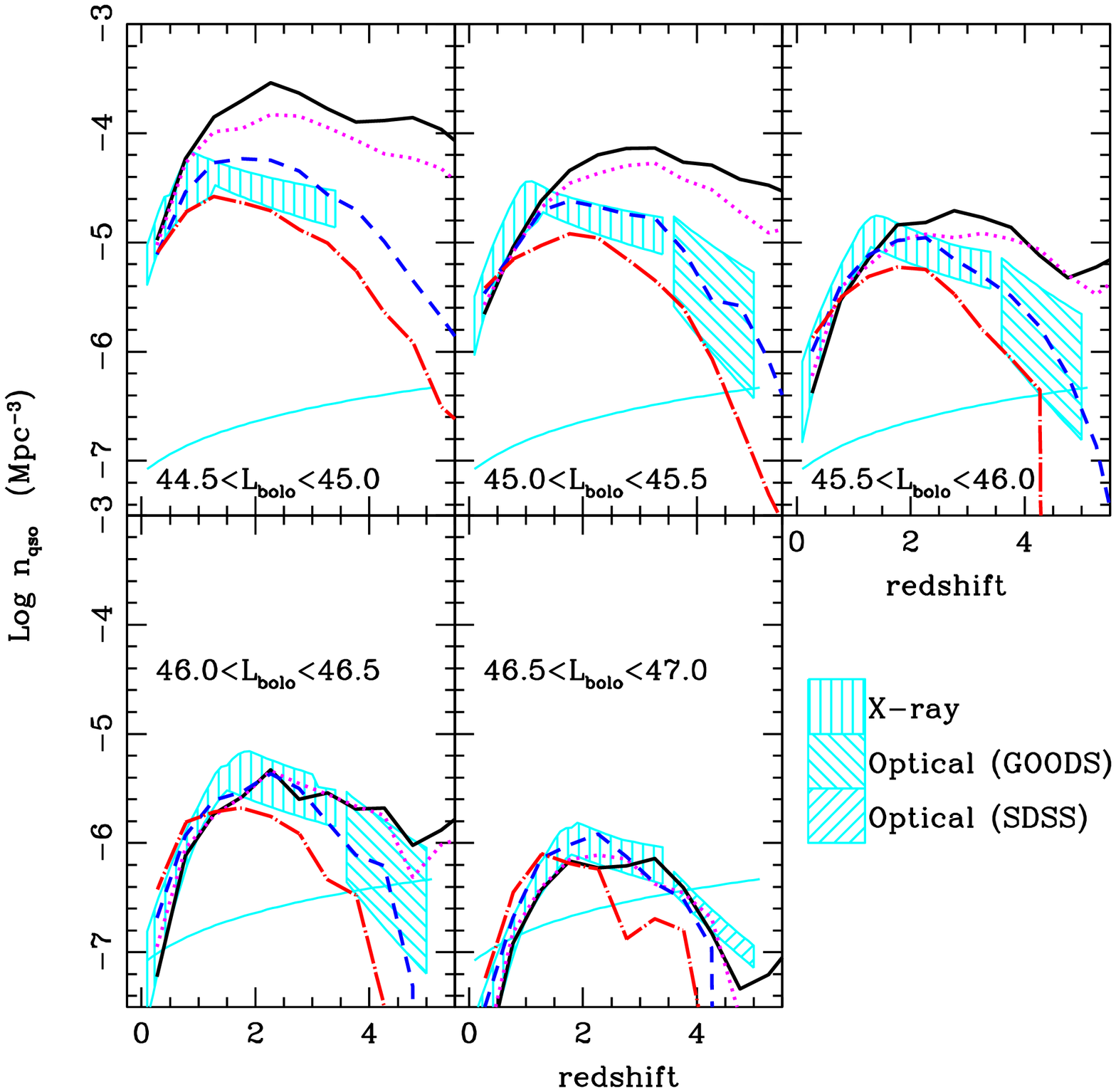}{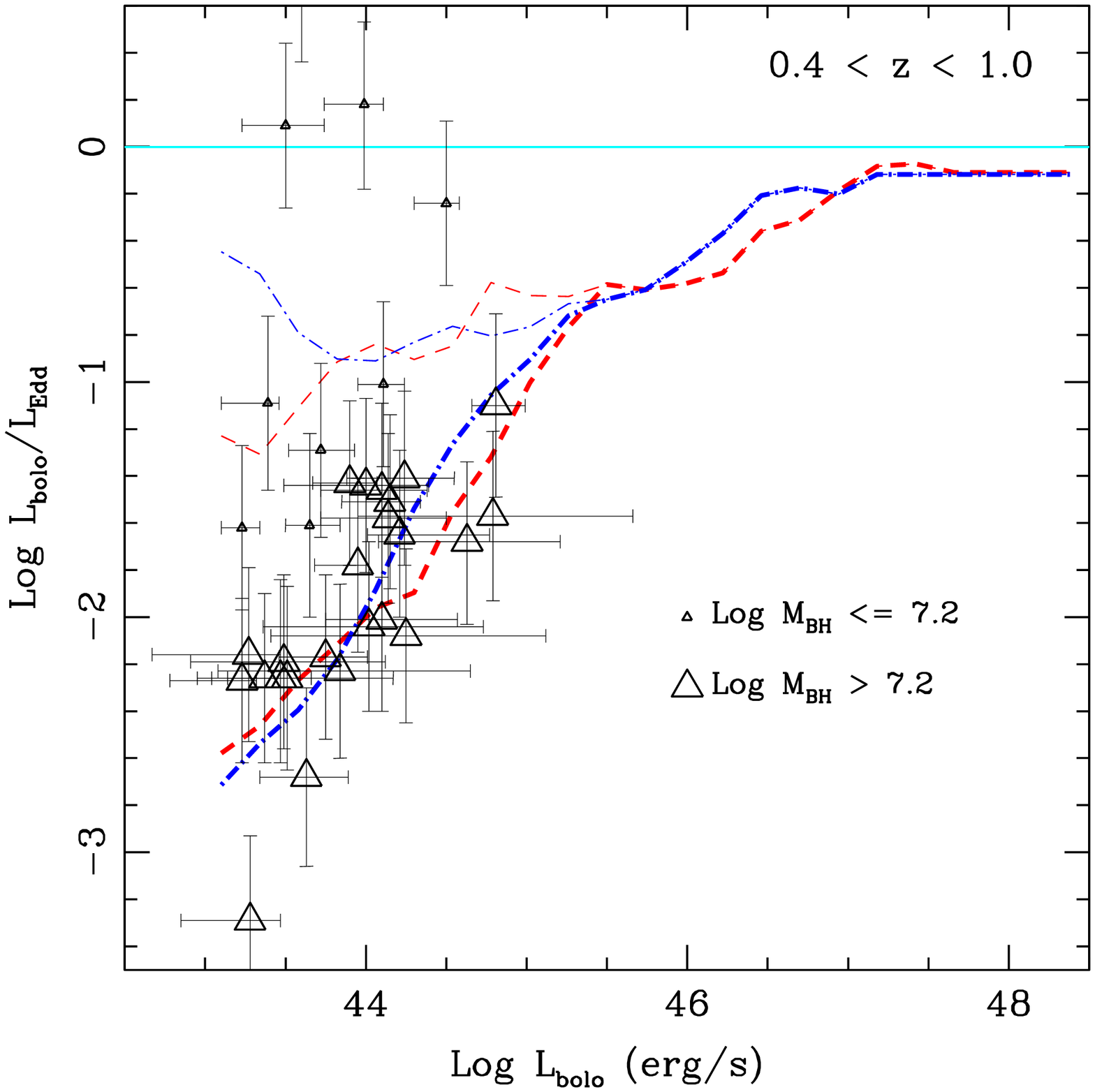}
\caption{Left panel: evolution of the QSO number density predicted by
{\sc morgana} as a function of $\sigma_0$ (equation 1).  The
black/solid, magenta/dotted, blue/dashed and red/dot-dashed lines
refer respectively to $\sigma_0=0$, 30, 60 and 90 km s$^{-1}$.  The
shaded areas give the range allowed by observations, as estimated from
X-ray (Ueda et al. 2003; Barger et al. 2005; La Franca et al. 2005)
and optical (SDSS, Fan et al. 2003; GOODS, Fontanot et al. 2006a).
Right panel: predicted Eddington ratios as a function of bolometric
luminosity for accreting black holes more massive than $10^{7.2}$
M$_\odot$ (thick lines) and for all blacks (thin lines).  Data are
from Ballo et al. (2006), large points refer to black holes more
massive than $10^{7.2}$ M$_\odot$, smaller points to smaller black
holes.}
\label{kinfeed}
\end{figure}

In Fontanot et al. (2006b) we show that kinetic feedback is a very
good candidate to be the main physical cause of the observed
downsizing of AGNs.  The left panel of figure~\ref{kinfeed} shows the
predicted number density of AGNs in bins of bolometric luminosity ad
as a function of redshift; the lines show models with increasing
values of $\sigma_0$ (equation 1).  The effect of kinetic feedback is
that of suppressing accretion onto black holes (by ejecting gas out of
the galaxy) in small, gas-rich bulges at high redshift, thus moving
the peak of low level activity to lower redshift, while the brighter
quasars, associated to more massive bulges, are hardly influenced.
This also implies that the low-level activity that gives rise to the
bulk of the hard X-ray background is due to relatively large black
holes hosted in middle-sized bulges.  This prediction is indeed in
line with the recent findings of Ballo et al. (2006) who, analyzing a
sample of X-ray active galaxies at $0.4<z<1$ in the GOODS fields,
computed bolometric luminosities from optical (nuclear) and X-ray
data, and inferred black hole masses from bulge masses. The agreement
between model predictions and data, shown in figure~\ref{kinfeed},
gives support to this scenario.  However, a closer look to the figure
reveals that the agreement gets worse when the smaller black holes are
concerned.  Is there a problem with small bulges/black holes?

\section{Something missing?}

\begin{figure}
\plottwo{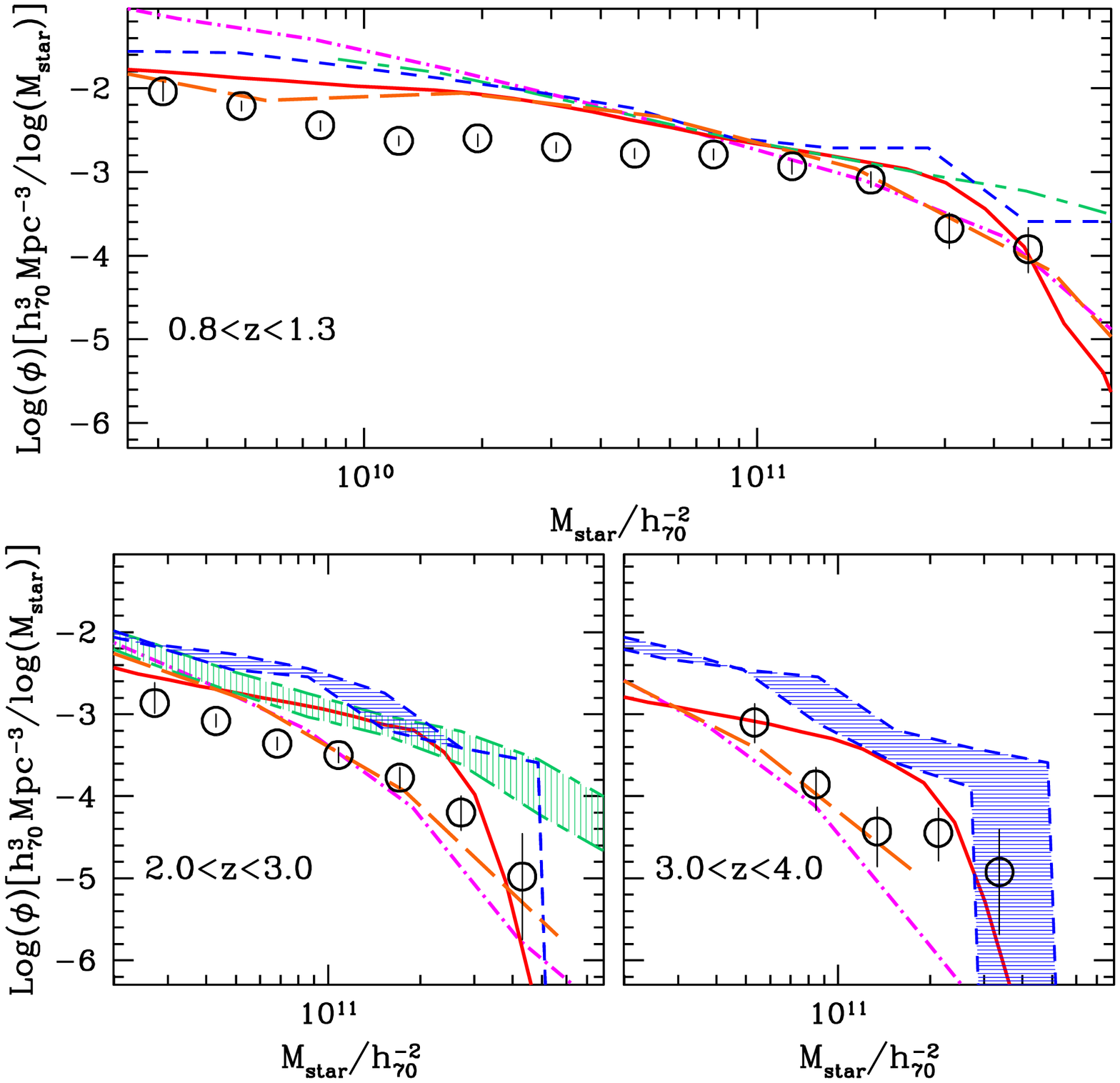}{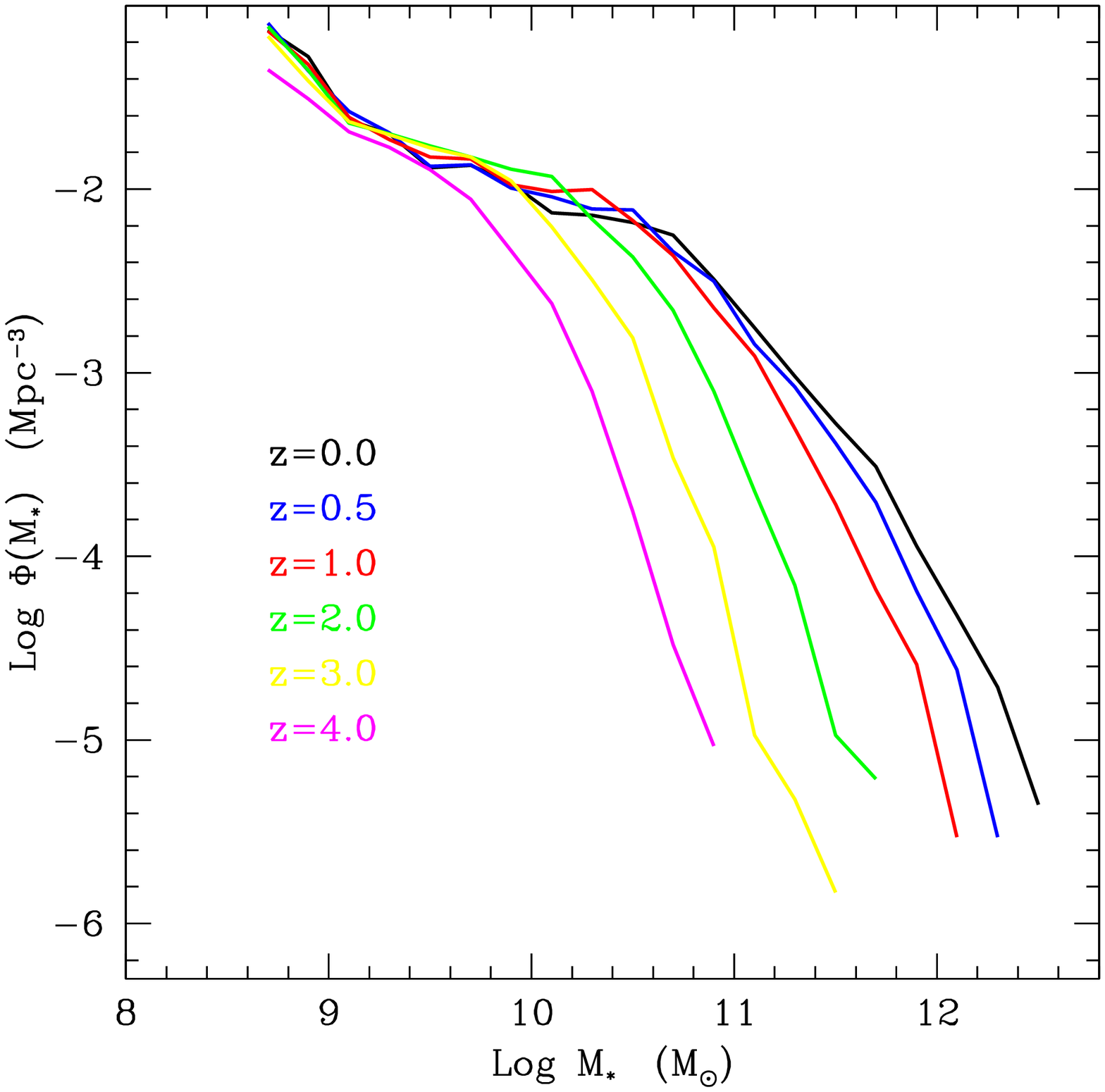}
\caption{Left panel: observed stellar mass function (GOODS-MUSIC
sample; Fontana et al. 2006) compared to the predictions of Bower et
al. (2006; red lines), Menci et al. (2006; dot-dashed purple lines),
{\sc morgana} (yellow long-dashed lines), Nagamine et al. (2005a, blue
dashed lines or areas), Nagamine et al. (2005b; green
long-short-dashed lines or areas).  Right panel: evolution of the
stellar mass function according to {\sc morgana}.}
\label{stellarmf}
\end{figure}

Despite the deep differences in the implementation, several galaxy
formation models are now giving consistent results.  It is then very
interesting to examine those cases where the results are consistently
discrepant with observations.  One result in this sense has recently
been obtained with the GOODS-MUSIC sample (Grazian et al. 2006), which
exploits the imaging performed with HST/ACS in the optical, VLT in the
U and NIR and Spitzer/IRAC in the MIR to give robust estimates of
photometric redshifts (trained on a large basis of available
spectroscopic redshifts) and stellar masses for a large sample of
galaxies with $K_s<23.5$.  Figure~\ref{stellarmf}, left panel, shows
the stellar mass function of Fontana et al. (2006), compared to {\sc
morgana} and to the models of Bower et al. (2006), Menci et al. (2006)
and the N-body+hydro codes of Nagamine et al. (2005a,b).  While the
build-up of massive galaxies is roughly followed, all models
consistently overpredict at $z\sim1$ the number of galaxies with
stellar masses $\sim10^{10}$ M$_\odot$.

More than a simple disagreement by a modest factor of two, this
discrepancy points to a significantly different evolution of the
stellar mass function in the reality and in the model: the real
stellar mass function evolves first in the low-mass tail, while, as
shown in the right panel of figure~\ref{stellarmf}, the model mass
function is remarkably constant in the same mass range.  If we now
define ``downsizing'' as the tendency of smaller galaxies to increase
significantly in number at low redshift, at variance with the
relatively stable number density of the more massive galaxies, then
this discrepancy points to an insufficient degree of downsizing in
the models.

{\sc morgana} has been written in order to easily introduce new
physical processes.  We have exploited this feature to insert many
recipes to suppress the number of such objects, without success.  As
AGN feedback is unlikely to be in play in these small galaxies, which
were shown above to host slowly accreting black holes, this
discrepancy strongly hints that there is some form of feedback that we
still do not understand.  A deeper study of ``gastrophysics'' is
needed to make sense of this basic observation.

\acknowledgements 

This work has been developed in collaboration with Fabio Fontanot,
Giuliano Taffoni, Stefano Cristiani, Paolo Tozzi, Laura Silva, Lucia
Ballo, Andrea Grazian, Adriano Fontana and the GOODS teams of Trieste
and Roma.

\end{document}